\documentclass[12pt]{article}   
\newcommand{\half}{\frac12}

\newcommand{\beq}{\begin{eqnarray}}
\newcommand{\eeq}{\end{eqnarray}}

\def \bi{\bibitem}
\def\){\right)}
\def\({\left( }
\def\]{\right] }
\def\[{\left[ }

\def\tr{{\rm\ Tr}}

\def\eps{\epsilon}

\newcommand{\vev}[1]{\langle#1\rangle}
\def\be{\begin{equation}}
\def\ee{\end{equation}}
\def\bea{\begin{eqnarray}}
\def\eea{\end{eqnarray}}
\newcommand{\eref}[1]{(\ref{#1})}

\newcommand{\rem}[1]{}
\def\tr{{\rm tr}}
\def\half{{1\over 2}}
\def\NN{{\cal N}}

\def\none{$\NN=1$}
\def\ntwo{$\NN=2$}

\def\nfour{$\NN=4$}

\def\rarr{\rightarrow}

\def\ZZ{{\bf Z}}

\def\ltap{\ \raise.3ex\hbox{$<$\kern-.75em\lower1ex\hbox{$\sim$}}\ }
\def\gtap{\ \raise.3ex\hbox{$>$\kern-.75em\lower1ex\hbox{$\sim$}}\ }

\title{Supergravity and  a Confining Gauge Theory: \\
Duality Cascades\\ and \\ $\chi$SB--Resolution of Naked Singularities}
\author{Igor R. Klebanov$^a$ and Matthew J. Strassler$^b$\\ \\
\small \sl $^a$Department of Physics, Princeton University\\
         \small \sl Princeton, NJ 08544, USA\\ 
  \small \sl $^b$School of Natural Sciences, Institute for Advanced Studies,\\
\small \sl Princeton, NJ 08540, USA\\ \\
  }
\begin{document}
\setlength{\baselineskip}{16pt}
\begin{titlepage}
\maketitle
\begin{picture}(0,0)(0,0)
\put(300,325){PUPT-1944}
\put(300,340){IASSNS--HEP--00/56}
\end{picture}
\vspace{-36pt}
\begin{abstract}
We revisit the singular IIB supergravity solution describing $M$
fractional 3-branes on the conifold [hep-th/0002159].  Its 5-form flux
decreases, which we explain by showing that the relevant
${\cal N}=1$ SUSY $SU(N+M)\times SU(N)$ gauge theory undergoes
repeated Seiberg-duality transformations in which $N\rightarrow N-M$.
Far in the IR the gauge theory confines; its chiral symmetry breaking
removes the singularity of hep-th/0002159 by deforming the conifold.
We propose a non-singular pure-supergravity background dual to the
field theory on all scales, with small curvature everywhere if the `t
Hooft coupling $g_s M$ is large. In the UV it approaches that of
hep-th/0002159, incorporating the logarithmic flow of couplings.  In
the IR the deformation of the conifold gives a geometrical realization
of chiral symmetry breaking and confinement.  We suggest that pure
\none\ Yang-Mills may be dual to strings propagating at small $g_sM$
on a warped deformed conifold.  We note also that the standard model
itself may lie at the base of a duality cascade.
\end{abstract}
\thispagestyle{empty}
\setcounter{page}{0}
\end{titlepage}

\section{Introduction}

A fruitful extension of the basic AdS/CFT correspondence 
\cite{jthroat,US,EW} stems
from studying branes at conical singularities \cite{ks,Kehag,KW,MP}.
Consider, for instance, a stack of D3-branes placed at the apex of a 
Ricci-flat 6-d cone $Y_6$ whose base is a 5-d Einstein manifold
$X_5$. Comparing the metric with the D-brane description leads
one to conjecture that type IIB string theory on $AdS_5\times X_5$
is dual to the low-energy limit of the world volume theory on
the D3-branes at the singularity.

A useful example of this correspondence has been to study D3-branes on
the conifold \cite{KW}.  When the branes are placed at the
singularity, the resulting ${\cal N}=1$ superconformal field theory
has gauge group $SU(N)\times SU(N)$.  It contains chiral superfields
$A_1, A_2$ transforming as $({\bf N},{\bf\overline N})$ and
superfields $B_1,B_2$ transforming as $({\bf\overline N},{\bf N})$,
with superpotential ${\cal W} =\lambda \epsilon^{ij} \epsilon^{kl}{\rm
Tr} A_iB_kA_jB_l$.  The two gauge couplings do not flow, and indeed
can be varied continuously without ruining conformal invariance.

For many singular spaces $Y_6$ there are also fractional D3-branes
which can exist only within the singularity \cite{gipol,doug,GK,KN}.  
These fractional D3-branes are D5-branes wrapped over (collapsed)
2-cycles at the singularity.
In the case of the
conifold, the singularity is a point.  The addition of $M$ fractional
branes at the singular point changes the gauge group to $SU(N+M)\times
SU(N)$; the four chiral superfields remain, now in the representation
$({\bf N+M},{\bf\overline N})$ and its conjugate, 
as does the superpotential \cite{GK,KN}.
The theory is no longer conformal.  Instead, the relative gauge
coupling $g_1^{-2}-g_2^{-2}$ runs logarithmically, as pointed out in
\cite{KN}, where the supergravity equations corresponding to this
situation were solved to leading order in $M/N$.  In \cite{KT} this
solution was completed to all orders; the conifold suffers logarithmic
warping, and the relative gauge coupling runs logarithmically at all
scales.  The D3-brane charge, i.e. the 5-form flux,
decreases logarithmically as well.
However, the logarithm in the solution is not cut off at small
radius; the D3-brane charge eventually 
becomes negative and the metric becomes
singular.

In \cite{KT} it was conjectured that this solution corresponds to a
flow in which the gauge group factors repeatedly drop in size by $M$
units, until finally the gauge groups are perhaps $SU(2M)\times SU(M)$
or simply $SU(M)$.  It was further suggested that the strong dynamics
of this gauge theory would resolve the naked singularity in the
metric.  Here, we show that this conjecture is correct.  The flow is
in fact an infinite series of Seiberg duality transformations --- a
``duality cascade'' --- in which the number of colors repeatedly drops
by $M$ units.  Once the number of colors in the smaller gauge group is
fewer than $M$, non-perturbative effects become essential.  We will
show that these gauge theories have an exact anomaly-free $\ZZ_{2M}$
R-symmetry, which is broken dynamically, as in pure \none\ Yang-Mills
theory, to $\ZZ_2$.  In the supergravity, this occurs through the
deformation of the conifold.\footnote{For a five-dimensional
supergravity approach to chiral symmetry breaking, see \cite{DZ}.}  In
short, the resolution of the naked singularity found in \cite{KT}
occurs through the chiral symmetry breaking of the gauge theory. The
resulting space, {\it a warped deformed conifold}, is completely
nonsingular and without a horizon, leading to confinement.  If the
low-energy gauge theory has fundamental matter, a horizon appears
and leads to screening.

\section{Branes and Fractional Branes on the Conifold}

\subsection{The Conifold}

The conifold is described by the following equation in
${\bf C}^4$:
\begin{equation} \label{conifoldA}
\sum_{n=1}^4 z_n^2 = 0\ .
\end{equation}
Equivalently, using $z_{ij} ={1\over \sqrt{2}}\sum_n \sigma^n_{ij} z_n$, where $\sigma^n$ are
the Pauli matrices for $n=1,2,3$ and $\sigma^4$ is $i$ times
the unit matrix, it may be written as
\begin{equation} \label{conifold}
\det_{i,j} z_{ij} = 0\ .
\end{equation}
This is a cone whose base is a coset space $T^{11}= (SU(2)\times
SU(2))/U(1)$, with topology $S^2\times S^3$ and symmetry group
$SU(2)\times SU(2)\times U(1)$.  As discussed in \cite{GK}, fractional
D3 branes at the singularity $z_n =0$ of the conifold are simply
D5-branes which are wrapped on the $S^2$ of $T^{11}$.  The Einstein
metric of $T^{11}$ may be written down explicitly \cite{cd}:
\begin{equation} \label{co}
d s_{T^{11}}^2=
{1\over 9} \bigg(d\psi + 
\sum_{i=1}^2 \cos \theta_i d\phi_i\bigg)^2+
{1\over 6} \sum_{i=1}^2 \left(
d\theta_i^2 + {\rm sin}^2\theta_i d\phi_i^2
 \right)
\ .
\end{equation}
It will be useful to employ
the following basis of 1-forms on the compact space
\cite{MT}:
\bea \label{fbasis}
g^1 = {e^1-e^3\over\sqrt 2}\ ,\qquad
g^2 = {e^2-e^4\over\sqrt 2}\ , \nonumber \\
g^3 = {e^1+e^3\over\sqrt 2}\ ,\qquad
g^4 = {e^2+ e^4\over\sqrt 2}\ , \nonumber \\
g^5 = e^5\ ,
\eea
where
\begin{eqnarray}
e^1\equiv - \sin\theta_1 d\phi_1 \ ,\qquad
e^2\equiv d\theta_1\ , \nonumber \\
e^3\equiv \cos\psi\sin\theta_2 d\phi_2-\sin\psi d\theta_2\ , \nonumber\\
e^4\equiv \sin\psi\sin\theta_2 d\phi_2+\cos\psi d\theta_2\ , \nonumber \\
e^5\equiv d\psi + \cos\theta_1 d\phi_1+ \cos\theta_2 d\phi_2 \ .
\end{eqnarray}
In terms of this basis, the Einstein metric on $T^{11}$ assumes the
form
\be
ds^2_{T^{11}}= {1\over 9} (g^5)^2 + {1\over 6}\sum_{i=1}^4 (g^i)^2
\ ,
\ee
and the metric on the conifold is
\be
ds_6^2 = dr^2 + r^2 ds^2_{T^{11}}
\ .
\ee

\subsection{The Gauge Theory}

If we place $N$ D3-branes and $M$ fractional D3-branes on
the conifold, we obtain an $SU(N+M)\times SU(N)$ gauge group.
The two gauge group factors have
holomorphic scales $\Lambda_1$ and $\tilde\Lambda_1$.  The
matter consists of two chiral superfields $A_1,A_2$ in the
$({\bf N+M,\overline{ N} })$ representation and two fields $B_1,B_2$ in the
$({\bf \overline{N+M},N})$ representation.  The superpotential of the
model is 
\begin{equation}
W=\lambda_1\tr\ (A_iB_jA_kB_\ell)\epsilon^{ik}\epsilon^{j\ell}\ .
\end{equation} 
The model has a
$SU(2)\times SU(2)\times U(1)$ global symmetry; the first (second) factor
rotates the flavor index of the $A_i$ $(B_i)$, while
the ``baryon'' $U(1)$ sends $A_i\to A_i e^{i\alpha}$, $B_i\to B_i
e^{-i\alpha}$.\footnote{The question of whether this $U(1)$ is
actually gauged is subtle.  We believe that it is
global, and arguments for this were given in \cite{KW,MP,KWnew}.}
There are also two spurious $U(1)$ transformations,
one an R-symmetry and one a simple axial symmetry, 
under which $\lambda_1$, $\Lambda_1$ and $\tilde\Lambda_1$
are generally not invariant.   The charges of the matter and the couplings
under the symmetries (excepting the $SU(2)$ flavor symmetries)
are given in Table 1.
Although $U(1)_A$ and $U(1)_R$ are anomalous, there is a discrete
$\ZZ_{2M}$ R-symmetry under which the theory is invariant.  In particular, 
if we let 
\begin{equation}
[A_i,B_j] \to [A_i,B_j] e^{2\pi i n/4M}, \ n = 1,2,\dots, 2M\ ,
\end{equation}
and rotate the gluinos by $e^{2\pi i n/2M}$, then the superpotential
rotates by $e^{2\pi i n/M}$ with $\lambda_1$, $\Lambda_1$ and
$\tilde\Lambda_1$ unchanged.

\vskip .2 in 
\begin{tabular}{|l|c|c|c|c|c|c|c|}
\hline
&&&&&&& \\ 
\hfil  & $SU(N_+)$ &$SU(N)$ &\hfil $SU(2)$ \hfil
 &\hfil $SU(2)$ \hfil &\hfil $U(1)_B$ \hfil &\hfil $U(1)_A$ \hfil
&\hfil $U(1)_R$ \hfil
\\&&&&&&& \\ \hline
&&&&&&&\\&&&&&&&\\ [-12pt]
$A_1,A_2$&${\bf N_+}$&${\bf \overline{N}}$&${\bf 2}$&${\bf 1}$
&${1\over 2 N_+ N}$&${1\over 2 N_+ N}$&$\half$\\ &&&&&&&\\
$B_1,B_2$&${\bf \overline{N_+}}$&${\bf N}$&${\bf 1}$&${\bf 2}$&
$-{1\over 2 N_+ N}$&${1\over 2 N_+ N}$&$\half$\\ &&&&&&&\\
$\Lambda_1^{3N_+-2N}$&&& & &$0$&${2\over N_+}$&$2M$\\ &&&&&&&\\
$\tilde\Lambda_1^{3N-2N_+}$&&& & &$0$&${2\over N}$&$-2M$\\ &&&&&&&\\
$\lambda_1$& &&& &$0$&$-{2\over N_+N}$&$0$\\ &&&&&&&\\
\hline
\end{tabular}
\vskip .2 in 
Table 1.
{\it Quantum numbers in the $SU(N+M)\times SU(N)$ model; we have
written $N_+\equiv N+M$ for concision.  }
\vskip .2 in

The classical field theory is well aware that it represents branes
moving on a conifold \cite{KW,MP}.  Let us consider the case
where the $A_i$ and $B_k$ have diagonal expectation values, $\vev{A_i}
= {\rm diag} (a_i^{(1)}, \cdots,a_i^{(N)})$, $\vev{B_i} = {\rm diag}
(b_i^{(1)}, \cdots,b_i^{(N)})$.  The F-term equations for a
supersymmetric vacuum
\begin{equation}
B_1A_iB_2-B_2A_iB_1 = 0 \ , \qquad \ A_1B_kA_2-A_2B_kA_1 = 0
\end{equation}
are automatically satisfied in this case, while the D-term equations
require $|a_1^{(r)}|^2+|a_2^{(r)}|^2-|b_1^{(r)}|^2-|b_2^{(r)}|^2 = 0$.
Along with the phases removed by the maximum abelian subgroup of the
gauge theory, the D-terms leave only $3N$ independent complex
variables.  Define $n_{ik}^{(r)} = a_i^{(r)} b_k^{(r)}$; then
the D-term and gauge invariance conditions are satisfied by
using the $n_{ik}^{(r)}$ as coordinates.  These $4N$ complex
coordinates satisfy the condition, for each $r$, 
\begin{equation} \label{braneoncon}
\det_{i,k}n_{ik}^{(r)}= 0 \ .
\end{equation}
This is the same as equation (\ref{conifold}).  Thus, for each
$r=1,\cdots, N$, the coordinates $n_{11}^{(r)}$, $n_{12}^{(r)}$,
$n_{21}^{(r)}$, $n_{22}^{(r)}$, are naturally thought of as the
position of a D3-brane moving on a conifold.

There are various combinations of the fields and parameters which are
invariant under the global symmetries.  One is
\begin{equation}
I_1\sim \lambda_1^{3M}{\tilde\Lambda_1^{3N-2(N+M)}\over\Lambda_1^{3(N+M)-2N}}
\ [\tr\ (A_iB_jA_kB_\ell\epsilon^{ik}\epsilon^{j\ell})]^{2M}
\end{equation} 
In addition, there are simple invariants such as 
\begin{equation}
R_1^{(1)}  = {\tr\ [A_iB_j]\tr[A_kB_\ell]\epsilon^{ik}\epsilon^{j\ell}
\over \tr\ (A_iB_jA_kB_\ell\epsilon^{ik}\epsilon^{j\ell})} \ ;
\end{equation}
there are many other similar invariants, in each of which the same
number of $A$ and $B$ fields appear in numerator and denominator but
with color and flavor indices contracted differently.  Finally
there is a constant invariant
\begin{equation}
J_1\equiv \lambda_1^{(N+M)+N}\Lambda_1^{3(N+M)-2N}\tilde\Lambda_1^{3N-2(N+M)}
\end{equation}
which plays the role of a dimensionless complex coupling analogous to $\tau$
in \nfour\ Yang-Mills.

The superpotential of the model will be renormalized and takes the
general form
\begin{equation}
W = \lambda_1\tr\ (A_iB_jA_kB_\ell)\epsilon^{ik}\epsilon^{j\ell} 
\ F_1(I_1,J_1,R_1^{(s)})
\end{equation}
where $F_1$ is a function which we will not fully determine.

\subsection{The conformal case: $M=0$}

If there are no fractional D3-branes, then the $U(1)_R$ is
anomaly-free, and the theory is superconformal (or if the couplings
$g_1,g_2,\lambda$ are chosen completely arbitrarily, it will flow
until it becomes conformal in the infrared.)  There are two
dimensionless global invariants $\lambda^2\Lambda_1\tilde\Lambda_1$,
the overall coupling $\tau_1+\tau_2$, and $\tilde\Lambda_1/\Lambda_1$,
the relative coupling $\tau_1-\tau_2$, which are built purely from the
parameters and may be chosen arbitrarily.  Thus \cite{kinstwo,emop,KW}
there are two exactly marginal operators in the theory which preserve
the continuous global symmetries.  (There are other marginal operators
which partially preserve these symmetries.)  This was the case studied in
\cite{KW}, where it was shown the supergravity dual of this field
theory is simply $AdS_5\times T^{11}$.

  In order to match the two couplings to the moduli of the type IIB
theory on $AdS_5\times T^{11}$, one notes that the integrals over the
$S^2$ of $T^{11}$ of the NS-NS and R-R 2-form potentials, $B_2$ and
$C_2$, are moduli. In particular, the two gauge couplings are
determined as follows \cite{KW,MP}:
\be
{1\over g_1^2} + {1\over g_2^2} \sim e^{-\phi}
\ ,
\ee
\be \label{couplediff}
{1\over g_1^2} - {1\over g_2^2} \sim e^{-\phi}\left [
\left(\int_{S^2} B_2\right) - 1/2 \right ]
\ ,
\ee
where $(\int_{S^2} B_2)$ is normalized in such a way that its period is
equal to $1$.\footnote{These equations are crucial for relating the SUGRA
background to the field theory beta functions when the
theory is generalized to $SU(N+M)\times SU(N)$ \cite{KN,KT}.}
The matching between the moduli is one of the simplest checks of
the duality. It is further possible to build a detailed correspondence
between various gauge invariant operators in the $SU(N)\times SU(N)$
gauge theory and modes of the type IIB theory on $AdS_5\times T^{11}$
\cite{KW,Gubser,Ceres}.

In \cite{Ur,DasM}, it was noted that there exists a type IIA
construction \cite{AH} which is T-dual to $N$ D3-branes at the
conifold. It involves two NS5-branes: one oriented in the $(12345)$
plane, and the other in the $(12389)$ plane. The coordinate $x^6$ is
compactified on a circle of circumference $l_6$, and there are $N$
$(1236)$ D4-branes wrapped around the circle.  If the NS5-branes were
parallel, then the low-energy field theory would be the ${\cal N}=2$
supersymmetric $SU(N)\times SU(N)$ gauge theory with bifundamental
matter (this type IIA configuration is T-dual to $N$ D3-branes at the
${\cal N}=2$ $\ZZ_2$ orbifold singularity). Turning on equal and
opposite masses for the two adjoint chiral superfields corresponds to
rotating one of the NS5-branes. Under this relevant deformation the
$\ZZ_2$ orbifold field theory flows to the ${\cal N}=1$ supersymmetric
conifold field theory \cite{KW,DasM}.

In terms of the type IIA brane construction, the two gauge couplings
are determined by the positions of the NS5-branes along the $x^6$ circle.
If one of the NS5-branes is located at $x_6=0$ and the other at
$x_6=a$, then \cite{DasM}
\be \label{IIAcplgs}
{1\over g_1^2} ={l_6 - a\over g_s}\ ,\qquad
{1\over g_2^2} ={a\over g_s}\ .
\ee
The couplings are equal when the NS5-branes are located diametrically
opposite each other (in the type IIB language this corresponds to
$ \int_{S^2} B_2 $ being equal to half of its period).
As the NS5-branes approach each other, one of the couplings becomes strong.
This simple geometrical picture will be useful for analyzing the RG
flows in the following sections.

\subsection{The RG cascade: $M>0$}

Now let us consider the effect of adding $M$ fractional D3-branes,
which as shown in \cite{GK} corresponds to wrapping
$M$ D5-branes over the $S^2$ of $T^{11}$.
The D5-branes serve as sources of the
magnetic RR 3-form flux through the $S^3$ of $T^{11}$. Therefore, the
supergravity dual of this field theory involves $M$ units of the 3-form
flux, in addition to $N$ units of the 5-form flux:
\be
\int_{S^3} F_3 = M\ ,\qquad\qquad \int_{T^{11}} F_5 = N
\ .
\ee
In the SUGRA description the 3-form flux is the source of conformal
symmetry breaking. Indeed, now $B_2$ cannot be kept constant and
acquires a radial dependence \cite{KN}:
\be \label{gravrun}
\int_{S^2} B_2 \sim M e^{\phi} \ln(r/r_0)
\ ,
\ee
while the dilaton stays constant at least to linear order in $M$.
Since the $AdS_5$ radial coordinate $r$ is dual to the RG scale 
\cite{jthroat,US,EW},
(\ref{couplediff}) implies a logarithmic running of
${1\over g_1^2} - {1\over g_2^2}$ in the $SU(N+M)\times SU(N)$ gauge theory.
This is in accord with the exact $\beta$-functions:
\beq \label{SVexact} 
&{d\over d {\rm log} (\Lambda/\mu)}
{8 \pi^2\over g_1^2} & \sim 3(N +M) - 2N (1- \gamma)\ ,\\
&{d\over d {\rm log} (\Lambda/\mu)}
{8 \pi^2 \over g_2^2} & \sim 3N - 2(N+M) (1-  \gamma)\ ,
\eeq
where $\gamma$ is the anomalous dimension of operators ${\rm Tr} A_i B_j$.
A priori, the conformal invariance of the field theory for $M=0$
requires that $\gamma =-{1\over 2} + O(M/N)$.
Taking the difference of the two equations in (\ref{SVexact}) we then
find
\be \label{betafun} 
{8\pi^2 \over g_1^2} - {8\pi^2 \over g_2^2} \sim
 M \ln (\Lambda/\mu) [ 3 + 2 
(1-\gamma)]\ ,
\ee
in agreement with (\ref{gravrun}) found on the SUGRA side.  The
constancy of the dilaton $\phi$ to order $M$ is consistent with the
field theory only if $\gamma =-{1\over 2} + O[(M/N)^2]$.  Fortunately,
the field theory in Table 1 has an obvious symmetry $M\to -M, N\to
N+M$, which to leading order in $M/N$ is $M\to -M$ with $N$ fixed.
Clearly $\gamma$ is even under this symmetry and so cannot depend on
$M/N$ at first order.

The SUGRA analysis of \cite{KN} was carried out to the linear order in
$M/N$. Luckily, it is possible to construct an exact solution taking
into account the back-reaction of $H_3$ and $F_3$ on other fields
\cite{KT}. In this solution $e^\phi = g_s$ is exactly constant, which
translates into the vanishing of the $\beta$-function for ${1\over
g_1^2} + {1\over g_2^2}$ in the dual field theory. 
As in \cite{KN}, \footnote{We are not keeping track of the overall
factor multiplying $M$, which is determined by the flux quantization.}
\be \label{closedf}
F_3 = M\omega_3\ ,  \qquad\qquad 
B_2 = 3 g_s M \omega_2 \ln (r/r_0)
\ ,
\ee
\be
H_3 = dB_2 = 3 g_s M {1\over r} dr\wedge \omega_2\ ,
\ee
where 
\be
\omega_2 = {1\over 2}(g^1\wedge g^2 + g^3 \wedge g^4)= 
{1\over 2} (\sin\theta_1 d\theta_1 \wedge d\phi_1- 
\sin\theta_2 d\theta_2 \wedge d\phi_2 )\ ,
\ee
\be
\omega_3 = {1\over 2} g^5\wedge (g^1\wedge g^2 + g^3 \wedge g^4)\ .
\ee
The relative factor of $3$ in (\ref{closedf}), which is related to the 
coefficients in the metric (\ref{co}), 
appears to be related to the factor of $3$
in the ${\cal N}=1$ beta function (\ref{betafun}). This gives the
correct value of beta function from a purely geometrical point of view. 

Both $\omega_2$ and $\omega_3$ are closed.
Note also that
\be \label{duality}
g_s \star_6 F_3 = - H_3\ ,\qquad g_s F_3 =  \star_6 H_3\ ,
\ee
where $\star_6$ is the Hodge dual with respect to the metric
$ds_6^2$. Thus, the complex 3-form $G_3$
satisfies the self-duality condition
\be
\star_6 G_3 = i G_3\ , \qquad\qquad G_3 = F_3 + {i\over g_s} H_3\ .
\ee
This is consistent with $G_3$ being either a $(0,3)$ form or a
$(2,1)$ form on the conifold. The Calabi-Yau form carries $U(1)_R$ charge
equal to 2, while $G_3$ does not transform under the $U(1)_R$.
Hence, the only consistent possibility appears to be
that $G_3$ is a harmonic $(2,1)$ form.\footnote{We are grateful to
S. Gubser and E. Witten for discussions on this issue.}

It follows from (\ref{duality}) that
\be \label{dilcons}
g_s^2 F_3^2 = H_3^2
\ ,
\ee
which implies that the dilaton is constant, $\phi=0$.
Since $F_{3\mu\nu\lambda} H_3^{\mu\nu\lambda} =0$, the RR scalar
vanishes as well.
The 10-d metric is
\be \label{fulsol}
ds^2_{10} =   h^{-1/2}(r)   dx_n dx_n 
 +  h^{1/2} (r)  (dr^2 + r^2 ds^2_{T^{11}} )\ ,
\ee
where
\be \label{nonharm}
h(r) = b_0 + 4\pi {g_s N + a (g_s M)^2 \ln (r/r_0) + a (g_s M)^2/4
\over r^4}
\ee
and $a$ is a constant of order 1.
Note that, for the ansatz (\ref{fulsol}), the solution 
for $h$ may be determined
from the trace of the Einstein equation:\footnote{
We are grateful to A. Tseytlin for explaining this to us.}
\be
h^{-3/2} \nabla^2_6 h \sim g_s^2 F_3^2 + H_3^2 = 2 g_s^2 F_3^2 
\ ,
\ee
where $\nabla^2_6$ is the Laplacian on the conifold. Since 
$F_3^2 \sim M^2 r^{-6} h^{-3/2}$, 
the solution (\ref{nonharm}) follows directly. 

An important feature of this background, which is not visible to linear
order in $M$, is that $F_5$ acquires a radial dependence \cite{KT}.
This is because
\be
F_5 = dC_4 + B_2\wedge F_3\ ,
\ee
and $\omega_2\wedge \omega_3 \sim {\rm vol}(T^{11})$.
Thus, we may write
\be
F_5 = {\cal F}_5 + *{\cal F}_5 \ , \qquad  {\cal F}_5
={\cal K}(r) {\rm vol}({\rm T}^{11})
\ ,
\ee
and 
\be
{\cal K} (r) = N + a g_s M^2 \ln (r/r_0)
\ .
\ee
The novel phenomenon in this solution is that the 5-form flux present
at the UV scale $r=r_0$ may completely disappear by the time we
reach a scale $r=\tilde r$ where ${\cal K}(\tilde r)=0$. This is related to
the fact that the flux
$\int_{S^2} B_2$ is not a periodic variable in the SUGRA solution:
as this flux goes through a period, ${\cal K}(r) \rightarrow 
{\cal K}(r) - M$
which has the effect of decreasing the 5-form flux by $M$ units.
We will shortly relate this decrease, which we refer to for now
as the ``RG cascade'', to Seiberg duality.

In order to eliminate the asymptotically flat region for large $r$
we use the well-known device of setting $b_0=0$ (this corresponds
to choosing the special solution of sec. 5 in \cite{KT}).
In terms of the scale $\tilde r$, we then have
\be {\cal K}(r) = a g_s M^2\ln (r/\tilde r)\ ,\qquad 
h(r) = {4\pi g_s\over r^4} [{\cal K}(r)+ a g_s M^2/4]
\ee
This solution has a naked singularity at $r=r_s$ where $h(r_s)=0$.
Writing
\be \label{UVs}
h(r) = {L^4\over r^4} \ln (r/r_s)\ , \qquad L^2 \sim  g_s M
\ ,
\ee
we then have a purely logarithmic RG cascade:
\be \label{dualmet}
ds^2 = {r^2\over L^2 \sqrt{\ln (r/r_s)} } dx_n dx_n
+ { L^2 \sqrt{\ln (r/r_s)}\over r^2} dr^2 + L^2 \sqrt{\ln (r/r_s)} 
ds^2_{T^{11}}\ .
\ee
This is essentially the metric of sec. 5 in \cite{KT} expressed in
terms of a different radial coordinate.  Since $T^{11}$ expands
slowly toward large $r$, the curvatures decrease there so
that corrections to the SUGRA are negligible. Therefore, there is no
obstacle for using this solution as $r\rightarrow \infty$ where the
5-form flux diverges.  The field theory explanation of the divergence
is that the RG cascade goes on forever as the scale is increased,
generating bigger and bigger $N$ in the UV.

As the theory flows to the IR, the cascade must stop, however, because
negative $N$ is physically nonsensical. Thus, we should not
be able to continue the solution (\ref{dualmet}) to $r< \tilde r$
where ${\cal K}(r)$ is negative. The radius of $T^{11}$ at $r=\tilde r$
is of order $\sqrt{g_s M}$. The gauge group at this scale is 
essentially $SU(M)$, and it is satisfying to see the appearance of
$g_s M$, which is the `t Hooft coupling. As usual, if the `t Hooft
coupling is large then the SUGRA solution has small curvatures.
Nevertheless, the fact that the solution of \cite{KT} is singular
tells us that it has to be modified, at least in the IR.
After understanding the RG cascade, we will study the dynamics
of the corresponding field theory, and will see how this
singularity is removed.

\section{The \none\ RG Cascade is a Duality Cascade}

We now trace the jumps in the rank of the gauge group to a well-known
phenomenon in the dual ${\cal N}=1$ field theory, namely, Seiberg
duality \cite{NAD}. The essential observation is that $1/g_1^2$ and
$1/g_2^2$ flow in opposite directions and, according to
(\ref{SVexact}), there is a scale where the $SU(N+M)$ coupling, $g_1$,
diverges. To continue past this infinite coupling, we perform a ${\cal
N}=1$ duality transformation on this gauge group factor.  The
$SU(N+M)$ gauge factor has $2N$ flavors in the fundamental
representation.  Under a Seiberg duality transformation, this becomes
an $SU(2N-[N+M]) = SU(N-M)$ gauge group with $2N$ flavors, which we
may call $a_i$ and $b_i$, along with ``meson'' bilinears $M_{ij}=
A_iB_j$.  The fields $a_i$ and $b_i$ are fundamentals and
antifundamentals of $SU(N)$, while the mesons are in the
adjoint-plus-singlet of $SU(N)$.  The superpotential after
the transformation
\begin{equation}
W = \lambda_1 \tr\ M_{ij}M_{k\ell}\eps^{ik}\eps^{j\ell}F_1(I_1,J_1,R_1^{(s)}) 
+ {1\over \mu} \tr\  M_{ij} a_ib_j \ ,
\end{equation}
where $\mu$ is the matching scale for the duality transformation
\cite{kinstwo}, shows the $M_{ij}$ are actually massive.  We may integrate
them out
\begin{equation}
0 = 2 \lambda_1 M_{k\ell}\eps^{ik}\eps^{j\ell}F_1(I_1,J_1,R_1^{(s)}) 
- {1\over \mu} \tr\ a_ib_j 
\end{equation}
leaving a superpotential
\begin{equation}
W = \lambda_2
\tr\ a_i b_j a_k b_\ell \eps^{ik}\eps^{j\ell} F_2(I_2,
J_2,R_2^{(s)}) 
\end{equation}
Here $F_2$, $\lambda_2$, $ I_2$, $J_2$ and $R_2$ are defined
similarly as in the original theory.  Thus we obtain an $SU(N)\times SU(N-M)$
theory which resembles closely the theory we started with.
\footnote{
The fact that the quartic superpotential is left roughly invariant by the
duality transformation in theories of this type has long been
considered of interest.  It was first noted in \cite{kinstwo}, where
it was used to study duality in $SO(3)$ gauge theories, and in
\cite{emop}, where its wider significance in Seiberg duality
transformations was established.}

Let us study the matching more carefully.  We define, for reasons
which will become clear in a moment, the strong coupling scale of the
$SU(N-M)$ factor to be $\tilde\Lambda_2$.  The strong coupling scale
of the $SU(N)$ factor is not the same as it was before the duality
(since the number of flavors in the $SU(N)$ gauge group has changed)
and its old scale $\tilde\Lambda_1$ must be replaced with a new strong
coupling scale $\Lambda_2$.  The matching conditions relating these
scales are of the form
\begin{equation}
\lambda_2\propto {1\over \mu^2\lambda_1}
\end{equation}
and
\begin{equation}
\Lambda_1^{3(N+M)-2N} \tilde\Lambda_2^{3(N-M)-2N}\propto \mu^{2N}\propto 
\lambda_1^{M}\tilde\Lambda_1^{3N-2(N+M)} \lambda_2^{-M}\Lambda_2^{3N-2(N-M)}
 \ .
\end{equation}
It is easy to check that
\begin{equation}
I_2\propto I_1 \ {\rm and}  \  J_2\propto 1/J_1 \ .
\end{equation}
(Note that the inversion of $J$ is a sign of electric-magnetic
duality, the generalization of $\tau\rarr -1/\tau$.)
Matching of baryon numbers in the Seiberg duality
assures that $(A)^{(N+M)}\sim (a)^{(N-M)}$.  We will not attempt
to match the $R_i^{(s)}$.

With these matchings, the dual theory has the global charges given in
Table 2.  Remarkably, this theory has the same form as the previous
one with $N\to N-M$.  Thus the renormalization group flow is
self-similar: the next step is that the $SU(N)$ gauge group now
becomes strongly coupled, and under a Seiberg duality transformation
the full gauge group becomes $SU(N-M)\times SU(N-2M)$, and so forth.
\vskip .2 in 
\begin{tabular}{|l|c|c|c|c|c|c|c|}
\hline
\hfil &$SU(N)$ & $SU(N_-)$  &\hfil $SU(2)$ \hfil
 &\hfil $SU(2)$ \hfil &\hfil $U(1)_B$ \hfil &\hfil $U(1)_A$ \hfil
&\hfil $U(1)_R$ \hfil
\\ \hline
&&&&&&&\\&&&&&&&\\ [-12pt]
$a_1,a_2$&${\bf \overline{ N}}$&${\bf N_-}$&${\bf 2}$&${\bf 1}$
&${1\over 2 NN_-}$&${1\over 2 NN_-}$&$\half$\\ &&&&&&&\\
$b_1,b_2$&${\bf N}$&${\bf \overline{N_-}}$&${\bf 1}$&${\bf 2}$&
$-{1\over 2 NN_-}$&${1\over 2 NN_-}$&$\half$\\ &&&&&&&\\
$\Lambda_2^{3N-2N_-}$&&& & &$0$&${2\over N}$&$2M$\\ &&&&&&&\\
$\tilde\Lambda_2^{3N_--2N}$&&& & &$0$&${2\over N_-}$&$-2M$\\ &&&&&&&\\
$\lambda_2$& &&& &$0$&$-{2\over NN_-}$&$0$\\ &&&&&&&\\
\hline
\end{tabular}
\vskip .2 in 
Table 2.
{\it Quantum numbers of the dual $SU(N)\times SU(N-M)$ theory; we have
written $N_-=N-M$ for concision.   }
\vskip .2 in

This flow will stop, of course, at or before the point where $N-kM$
becomes zero or negative.  Note that the Seiberg duality
transformation is the same in both the so-called conformal window
($3N_c>N_f>{3\over2}N_c$) and in the free magnetic phase
(${3\over2}\geq N_f>N_c+1$).  Even for $N_f=N_c+1$ the effect on the
superpotential described in \cite{nsexact} is not essential, since it
is accounted for in the function $F$.  The first significant changes
occur when $N_f=N_c$, since for $N_f\leq N_c$ the classical moduli
space is drastically modified.  Thus, the RG flow just described
proceeds step by step until the gauge group has the form
$SU(M+p)\times SU(p)$, where $0<p\leq M$.
At this point we should do a more careful analysis, which
we will carry out in the next section. 

It is instructive to consider the type IIA brane picture of the
duality cascade of $SU(N+M)\times SU(N)$ theories.\footnote{We should
remind the reader that the classical IIA picture of Seiberg duality,
introduced by \cite{EGK}, has the feature that it is not a string
duality but a motion which transforms one theory into its dual, one
which a priori need not leave the infrared physics invariant.  In
particular, there is no direct relation between the classical brane
motion, or semiclassical brane bending, and the actual dynamics of the
field theories. To see that the IIA story gives the right answer, one
should use its M theory generalization \cite{ewelliptic,HOO,ewmqcd,BIKSY}, but
even there, the M5-brane, about which only holomorphic information is
available, does not generally match the dynamics of the field theory,
which is not holomorphic.  These issues have been studied and
explained in detail; see especially \cite{horiMQCD}.  We mention this
only to warn our readers not to take this paragraph for more than a
heuristic argument.}  To implement such theories we have to add $M$
D4-branes stretched only one way between the D4-branes, rather than
all around the circle \cite{EGK,AH,ewelliptic,Dmfract}. Such D4-branes
are T-duals of the ``fractional'' branes which are the D5-branes
wrapped over the 2-cycle of $T^{11}$.  These new D4-branes violate the
balance of forces on the NS5-branes, and the latter undergo
logarithmic bending \cite{ewelliptic}, which is the RG
flow in this picture. Although the NS5 and NS5' brane bend along the
circular $x^6$ direction, this does not force them to intersect,
because they are oriented in perpendicular directions.  However, their
$x^6$ positions become equal somewhere away from the fractional
branes, and it seems natural to interpret this as a divergence of the
$SU(N+M)$ coupling.  Under such circumstances it is natural to move
the NS branes so as to eliminate this divergence, moving one of them
once around the $x^6$ circle.  When the NS and NS' brane cross during
this motion, the $N+M$ fractional D4-branes shrink to zero size and
then re-grow.  In doing so, they flip their orientation and become
anti-D4-branes.  Meanwhile, the other $N$ fractional branes stretch
more than once around the circle, but where they are doubled they are
partially cancelled by the $N+M$ anti-D4-branes.  This leaves $N-M$
D4-branes in one segment and $N$ in the other --- exactly our starting
point but with $N\to N-M$. After the crossing, the NS5-branes are
still bent in the same directions as before, so again their $x^6$
positions become equal and we are led to repeat the motion around the
circle.  Finally, the number $N$ becomes of order $M$ and something
more drastic should happen \cite{EGKRS}.  For this physics, the
analysis of \cite{HOO,ewmqcd,BIKSY} becomes essential.

\section{Chiral Symmetry Breaking and the Deformation of the Conifold}

The solution of \cite{KT} is well-behaved for large $r$ but becomes
singular at sufficiently small $r$. The solution must be modified in
such a way that this singularity is removed. In this section we argue
that the conifold (\ref{conifold}) should be replaced by the deformed
conifold
\begin{equation} \label{dconifold}
\sum_{i=1}^4 z_i^2 = -2\det_{i,j} z_{ij} = \epsilon^2\ ,
\end{equation}
in which the singularity of the conifold is removed
through the blowing-up of the  $S^3$ of $T^{11}$.

There are a number of arguments in favor of this idea.  One suggestive
observation is that in the solution of \cite{KT}, the source of the
singularity can be traced to the infinite energy in the $F_3$ field.
At all radii there are $M$ units of flux of $F_3$ piercing 
the $S^3$ of $T^{11}$, and when the $S^3$ shrinks to zero size
this causes $F_3^2$ to diverge.  If instead the $S^3$ remained
of finite size, as occurs in the deformed conifold, this problem would
be evaded.

However, the most powerful argument that the conifold is deformed
comes from the field theory analysis, which shows clearly that the
spacetime geometry is modified by the strong dynamics of the infrared
field theory.  We will see that the theory has a deformed moduli
space, with $M$ independent branches, each of which has the shape of a
deformed conifold.  The branches are permuted by the $\ZZ_{2M}$
R-symmetry, which is spontaneously broken down to $\ZZ_2$.  This
breaking of the R-symmetry is exactly what we would expect in a pure
$SU(M)$ \none\ Yang-Mills theory, although here it proceeds through
scalar as well as gluino expectation values.  The theory will 
also have domain walls, confinement, magnetic screening, and other
related phenomena.

The complete analysis of the nonperturbative dynamics of the field
theory in Table 1 is mathematically intensive, and we have not
attempted it.  In this section we present a simplified version of the
analysis which captures the physics which we are interested in.  In an
appendix we present more general (although still partial) results that
show our conclusions are robust.

In particular, our goal is to discover what happens in the far
infrared of the flow, where the D3 brane charge has cascaded (nearly)
to zero and only the $M$ fractional D3 branes remain.  If there are no
D3 branes left, we expect we have pure \none\ Yang-Mills in the far
infrared, a theory which breaks its $\ZZ_{2M}$ R-symmetry to $\ZZ_2$
and has $M$ isolated vacua, domain walls, and confinement.  However,
while this may be correct, we have no access to the supergravity
background through this analysis.  What we need is a probe which
can see if and how the fractional D3-branes have modified the
conifold itself.

The right choice, it turns out, is to probe the space with a single
additional D3 brane.  In this case the gauge group is $SU(M+1)\times
$``$SU(1)$'' --- in short, simply $SU(M+1)$ --- with fields $C_i$ and
$D_j$ in the ${\bf M+1}$ and ${\bf \overline{M+1}}$ representations,
$i,j=1,2$, and with superpotential $W = \lambda
C_iD_jC_kD_l\eps^{ik}\eps^{jl}$.  Define $N_{ij}=C_iD_j$, which is
gauge invariant.  As in the discussion surrounding equation
(\ref{braneoncon}), the expectation values of $N_{ij}$ specify the
position of the probe brane; in the classical theory, we have
$\det_{i,j} N_{ij}= 0$, indicating the probe is moving on the
original, singular conifold.  At low energy the theory can be written
in terms of these invariants and develops the
nonperturbative superpotential first written down by Affleck, Dine and
Seiberg \cite{adssup}
\begin{equation}
W_L = \lambda N_{ij}N_{k\ell} \eps^{ik}\eps^{j\ell} + (M-1)
\left[{2\Lambda^{3M+1}\over N_{ij}N_{k\ell} 
\eps^{ik}\eps^{j\ell}}\right]^{{1\over M-1}} \ .
\end{equation}
The equations for a supersymmetric
vacuum are
\begin{equation}
0 = \left(\lambda  -
\left[{2\Lambda^{3M+1}\over (N_{ij}N_{k\ell} 
\eps^{ik}\eps^{j\ell})^{M}}\right]^{{1\over M-1}} \right)N_{ij}\ .
\end{equation}
The apparent solution $N_{ij}=0$ for all $i,j$ actually gives infinity
on the right-hand side.  The only solutions are then
\begin{equation} 
(N_{ij}N_{k\ell} \eps^{ik}\eps^{j\ell})^{M} = 
{2\Lambda^{3M+1}\over \lambda^{M-1}} \ .
\end{equation}
As predicted, this equation has $M$ independent branches, in each of
which $ N_{ij}N_{k\ell} \eps^{ik}\eps^{j\ell}$ is a $M^{th}$ root of
${\Lambda^{3M+1}/ \lambda^{M-1}}$.  The $\ZZ_{2M}$ discrete
non-anomalous R-symmetry rotates $N_{ij}N_{k\ell} \eps^{ik}\eps^{j\ell}$ by a
phase $e^{2\pi i/M}$, and thus the $M$ branches transform into one
another under the symmetry.  In short, the $\ZZ_{2M}$ is spontaneously
broken down to $\ZZ_2$.  The low-energy effective
superpotential is
\begin{equation}
W = M\lambda \vev{N_{ij}N_{k\ell} \eps^{ik}\eps^{j\ell}}
\propto M\left[2\lambda\Lambda^{3M+1}\right]^{1/M}
\end{equation}
which reflects the $M$ branches. Most importantly,  on each of these branches
the classical condition on the $N_{ij}$ has been modified to read
\begin{equation} 
\det_{i,j} N_{ij}=  \half N_{ij}N_{k\ell} \eps^{ik}\eps^{j\ell} 
=\left({\Lambda^{3M+1}\over [2\lambda]^{M-1}}\right)^{1/M}
\end{equation}
Comparing with equation ({\ref{dconifold}) we see that the probe brane
in the quantum theory moves on the deformed conifold; the classical
singularity at the origin of the moduli space has been resolved
through chiral symmetry breaking.  

The above constraint on the expectation values for $N_{ij}$ implies
that in the perturbative region (where semiclassical analysis is
valid) they can break the gauge group only down to $SU(M)$, with no
massless charged matter.  This gauge theory is thus in the
universality class of pure $SU(M)$ Yang-Mills, and will share many of
its qualitative properties.  However, the existence of {\it massive}
matter $C_i,D_j$ in the fundamental representation of $SU(M)$ 
(note that if
$\vev{N_{11}}$ is large then $C_2,D_2$ have mass $\lambda N_{11}$)
implies that confinement occurs only in an intermediate range of
distances.  As in QCD with heavy quarks, pair production of the
massive quarks breaks the confining flux tubes, so a linear potential
between external sources exists only between the length scales
$1/\sqrt{T}$ and $m_q/T$, where $T$ is the string tension and $m_q$ is
the dynamical quark mass.  For $\vev{N_{11}}\sim \vev{N_{22}}\sim
\left({\Lambda^{3M+1}/ \lambda^{M-1}}\right)^{1/2M}$, their
minimal values, we expect little sign of a linear potential at any
length scale, as in physical QCD.  Only for $p=0$ do we expect
confinement at all scales.

More generally, for $1<p<M$, one obtains a moduli space
corresponding to $p$ probe branes moving on the deformed conifold. If
$p\ll M$, both the $SU(p)$ gauge coupling and its 't Hooft coupling
are small at the strong-dynamics scale of $SU(M+p)$.  Furthermore, the
$SU(p)$ factor has vanishing beta function in the far IR, where it has
three adjoint chiral superfields (namely, three of the $N_{ij}$) and
is essentially a copy of \nfour\ Yang-Mills. Consequently, we expect
no strong dynamics from the $SU(p)$ sector, and the theory is very
close to $SU(M+p)$ with $2p$ light flavors.  In this case a similar
analysis to the above is essentially correct. At large
expectation values, the gauge theory is broken to $SU(M)$ \none\
Yang-Mills times $SU(p)$ \nfour\ Yang-Mills, with massive states in
the bifundamental representation of the group factors.   Details
of this analysis are given in the appendix. As before,
pair production of these massive states eliminates confinement at
large distances; electric sources are screened by massive states which
leave them charged only under the nonconfining group $SU(p)$.

The pattern of chiral symmetry breaking gives us another qualitative
argument why the conifold must be deformed.  The original conifold has
a $U(1)_R$ symmetry under which the $z_{ij}$ in \eref{conifold} rotate
by a phase.  In Table 1 we saw this was broken by instantons to
$\ZZ_{2M}$, but for large $M$ this is a $1/M$ effect and need not show
up in the leading order supergravity.  However, if we expect the
infrared theory to behave similarly to pure \none\ Yang-Mills, then we
expect this symmetry to be spontaneously broken to $\ZZ_2$.  This
breaking is a leading-order effect and most definitely should be
visible in the supergravity.  The only natural modifications of the
conifold are its resolution and its deformation; only the latter
breaks the classical $U(1)_R$ symmetry, and it indeed breaks it to
$\ZZ_2$, as is obvious from equation \eref{dconifold}.

As a final argument, we consider expectations from the IIA/M brane
construction.  Classically we have NS and NS' branes filling
four-dimensional space and extending in the $v=x^4+ix^5$ and $
w=x^8+ix^9$ directions respectively.  They are separated along the
compact direction $x^6$ by a distance $a$, which along with
$l_6$ sets the two classical gauge couplings, as explained in \eref{IIAcplgs}.  In one $x^6$ segment between the NS and NS'
brane we suspend $M+1$ D4 branes; in the other there is only one D4
brane.  A single complete wrapped D4-brane --- our probe --- is free
to move anywhere in the $v,w,x^7$ space, independently of the other
branes, while the other $M$ suspended D4-branes are pinned to
$v=w=0$. To understand the quantum theory, we must move to M theory
\cite{HOO,ewmqcd,BIKSY}, where we combine $x^6$ with the new compact coordinate
$x^{10}$ using $t = e^{x^6+ix^{10}}$.  Classically the equations for
the NS and NS' brane are $w=0,t=1$ and $v=0,t=e^a$.  

The M theory expectation is that, in the quantum theory, the probe
brane will become an independent M5 brane wrapped on the $t$
directions, while the suspended D4-branes join with the NS and NS'
branes to make a single M5-brane, which we will refer to as our MQCD
brane.  This type of behavior was first seen in \ntwo\ and \none\
supersymmetric Yang-Mills \cite{ewelliptic,HOO,ewmqcd,BIKSY}.  Indeed the MQCD
brane which appears in our case should be very similar to that of
\none\ super-Yang-Mills, since in the limit the $x^6$ direction
becomes large they should become equal.  The brane for
super-Yang-Mills fills the coordinates $x^0,x^1,x^2,x^3$ and is
embedded in the coordinates $v=x^4+ix^5, w=x^8+ix^9, t =
e^{x^6+ix^{10}}$ as a Riemann surface defined through the equations
\begin{equation}\label{YMMbrane}
(vw)^M=\Lambda_L^{2M}\  , \ v^M = t \ .
\end{equation}
Notice classically the equations include $vw=0$, corresponding to
the presence of the NS and NS' brane.  However, the quantum Yang-Mills
M-brane has $vw$ equal to a nonzero constant, and has $M$ possible
orientations, one for each possible phase of a condensate.

What is the connection with the deformed conifold?  As shown in
\cite{Tdualdefcon}, a type IIA NS-brane and NS'-brane satisfying the
equation $vw=0$, that is, intersecting at a point, are T-dual to the
conifold.  This lifts without change to M theory.  We saw this
equation appears in the construction of classical Yang-Mills, and it
will appear in our classical theory as well.  Meanwhile, if the NS and
NS' branes are at the same $t$, that is, if they intersect, then they
can be deformed into a single object with equation $vw=$ constant $\neq
0$.  This object is T-dual to the {\it deformed} conifold.  Again this
also lifts without change to M theory.  Now notice that the Yang-Mills
M-brane has this as one of its defining equations \eref{YMMbrane}.
This shows the NS and NS' brane have been glued together into a single
object.  Without the suspended D4-branes, this could only occur if the
joined NS and NS'-brane had equal $t$ coordinates, but in the presence
of the suspended D4-branes, which extend along the $t$ direction, the
NS and NS'-branes can be separated in $t$, as in \eref{YMMbrane}.
Thus the Yang-Mills M-brane shows that the  suspended
D4-branes allow a quantum effect in M theory by which the conifold
can be deformed even when the two gauge couplings \eref{IIAcplgs} 
are both finite.

In our case, we similarly expect the two Riemann surfaces ---
the probe and the MQCD brane --- to have $M$
branches, with a continuous variable specifying the position of the
probe brane in the space, and a discrete variable labeling the
orientation of the MQCD brane.  However, when the probe is far away
and the $x^6$ direction is large, our MQCD brane should closely
resemble that of Yang-Mills.  We therefore expect the equations
governing it to have the same qualitative form.  In particular, we
expect that the $\ZZ_{2M}$ discrete symmetry rotating the phase of $t$
by $2\pi$ is broken to $\ZZ_2$, through the modification of the
equation $vw=0$ to $(vw)^M=$ constant.  By T-duality this indicates that the
classical conifold is quantum deformed by the fractional branes.

\section{Back to Supergravity: The Deformed Conifold Ansatz}

The field theory analysis of the previous section shows that
the naive $U(1)$ (really $\ZZ_{2M}$) R-symmetry is actually broken to a $\ZZ_2$.
On the other hand, the SUGRA background (\ref{fulsol}) has
an exact $U(1)$ symmetry realized as shifts of the angular coordinate 
$\psi$ on $T^{11}$. The presence of this unwanted symmetry in
the IR may also be the reason for the appearance of the 
naked singularity.

In this section we propose that the solution of this problem
is to replace the conifold by its deformation (\ref{dconifold})
in the ansatz (\ref{fulsol}). This indeed breaks the $U(1)$ symmetry
$z_k \rightarrow e^{i\alpha} z_k$, $k=1,\ldots, 4$, 
down to its $\ZZ_2$ subgroup $z_k \rightarrow - z_k$. Another reason
to focus on the deformed conifold is that it gives the correct
moduli space for the field theory, as shown in the previous section.

The metric of the deformed conifold was discussed in some detail in
\cite{cd,MT,Ohta}. It is diagonal in the basis (\ref{fbasis}):
\bea \label{metricd}
ds_6^2 = {1\over 2}\epsilon^{4/3} K(\tau)
\Bigg[ {1\over 3 K^3(\tau)} (d\tau^2 + (g^5)^2) 
 + 
\cosh^2 \left({\tau\over 2}\right) [(g^3)^2 + (g^4)^2]\nonumber \\
+ \sinh^2 \left({\tau\over 2}\right)  [(g^1)^2 + (g^2)^2] \Bigg]
\ ,
\eea
where
\be
K(\tau)= { (\sinh (2\tau) - 2\tau)^{1/3}\over 2^{1/3} \sinh \tau}
\ .
\ee
For large $\tau$ we may introduce another radial coordinate $r$ via
\be \label{changeofc}
r^3 \sim \epsilon^2 e^{\tau}\ ,
\ee
and in terms of this radial coordinate 
\be
ds_6^2 \rightarrow dr^2 + r^2 ds^2_{T^{11}}
\ .
\ee
The determinant of the metric (\ref{metricd}) is
\be
g_6 \sim \epsilon^8 \sinh^4 \tau\ ,
\ee
which vanishes at $\tau=0$.
Indeed, at $\tau=0$ the angular metric degenerates into
\be 
d\Omega_3^2= {1\over 2} \epsilon^{4/3} (2/3)^{1/3}
[ {1\over 2} (g^5)^2 + (g^3)^2 + (g^4)^2 ]
\ ,
\ee
which is the metric of a round $S^3$ \cite{cd,MT}.
The additional two directions, corresponding to the $S^2$ fibered
over the $S^3$, shrink as
\be {1\over 8} \epsilon^{4/3} (2/3)^{1/3}
\tau^2 [(g^1)^2 + (g^2)^2]
\ .\ee
In what follows we will set $\epsilon = 12^{1/4}$, so that
${1\over 2} \epsilon^{4/3} (2/3)^{1/3}=1$.

The collapse of the $S^2$ implies that at $\tau=0$ $F_3$
must lie within the remaining $S^3$,
\be F_3 (\tau=0) = M g^5\wedge g^3\wedge g^4\ ,
\ee
which may be shown to be a closed 3-form. On the other hand, for large
$\tau$, $F_3$ should approach its value
\be
{M\over 2} g^5 \wedge (g^1\wedge g^2 + g^3\wedge g^4)
\ee
found in the UV ansatz (\ref{closedf}). These two closed 3-forms differ
by an exact one, 
\be
g^5 \wedge (g^1\wedge g^2 - g^3\wedge g^4) = d( g^1\wedge g^3 + 
g^2\wedge g^4)
\ee
Therefore, the simplest ansatz which interpolates smoothly between
$\tau=0$ and large $\tau$ is
\bea
F_3 = M \left \{g^5\wedge g^3\wedge g^4 + d [ F(\tau) 
(g^1\wedge g^3 + g^2\wedge g^4)]\right \} \nonumber \\
= M \left \{g^5\wedge g^3\wedge g^4 (1- F))
+ g^5\wedge g^1\wedge g^2 F + F' d\tau\wedge
(g^1\wedge g^3 + g^2\wedge g^4) \right \}\ ,
\eea
with $F(0) = 0$ and $F(\infty)=1/2$.
Note also that this ansatz preserves the $\ZZ_2$ symmetry
which interchanges $(\theta_1,\phi_1)$ with $(\theta_2,\phi_2)$.

A similarly $\ZZ_2$-symmetric ansatz for $B_2$ is
\be
B_2 = g_s M [f(\tau) g^1\wedge g^2
+  k(\tau) g^3\wedge g^4 ]\ .
\ee
Using the identity
\be \label{nif}
g^5\wedge (g^1\wedge g^3 + g^2\wedge g^4) = -d (g^1\wedge g^2 - g^3\wedge
g^4)\ ,
\ee
we find that
\be
H_3 = dB_2 = g_s M [d\tau\wedge (f' g^1\wedge g^2
+  k' g^3\wedge g^4) + {1\over 2} (k-f) 
g^5\wedge (g^1\wedge g^3 + g^2\wedge g^4) ]\ .
\ee

We further have
\be
{\cal F}_5 = B_2\wedge F_3 = g_s M^2 \ell(\tau)
g^1\wedge g^2\wedge g^3\wedge g^4\wedge g^5\ ,
\ee
where
\be
\ell = f(1-F) + k F\ .
\ee

The most general radial ansatz for the 10-d metric, consistent
with the symmetries of the deformed conifold, is
\bea \label{genans}
ds_{10}^2 = A^2(\tau) dx_n dx_n + B^2(\tau) (d\tau^2) +
C^2(\tau) (g^5)^2 + D^2(\tau) [(g^3)^2 + (g^4)^2]
\nonumber \\
 + E^2(\tau)
[(g^1)^2 + (g^2)^2]\ .
\eea
The reason we are allowed to assume that $A,\ldots, E$ depend only
on $\tau$ is that before the introduction of the 3-form fields,
the metric has the form (\ref{genans}), and our ansatz for 
$F_3$ and $H_3$ does not break this symmetry. The flux of $F_3$
is distributed uniformly over the $S^3$ near the apex of the deformed
conifold; therefore, the $M$ D5 branes wrapped over the $S^2$ may
be thought of as smeared over the $S^3$.

It is not hard to check that $F_{3\mu\nu\lambda} H_3^{\mu\nu\lambda} =0$,
which implies that the RR scalar vanishes.
It is not a priori clear whether the dilaton is constant
for the deformed solution, but in what 
follows we will assume that such a background does exist, i.e.
that 
\be
\label{dilcon}
g_s^2 F_3^2 = H_3^2
\ .
\ee 
Furthermore,
guided by the simple form of the solution 
constructed in \cite{KT} and reviewed in section 2,
we will assume that the 10-d metric 
takes the following form:
\be \label{specans}
ds^2_{10} =   h^{-1/2}(\tau)   dx_n dx_n 
 +  h^{1/2}(\tau) ds_6^2 \ ,
\ee
where $ds_6^2$ is the metric of the deformed conifold (\ref{metricd}).
This is the same type of ``D-brane'' ansatz as (\ref{fulsol}), but with the
conifold replaced by the deformed conifold as 
the transverse space.  This form will also permit additional
D3-brane probes to be directly included in the ansatz.

The type IIB equations satisfied by the 3-form fields are
\be \label{threef}
d (e^{\phi}\star F_3) = F_5\wedge H_3\ ,\qquad 
d (e^{-\phi}\star H_3) = - g_s^2 F_5\wedge F_3\ .
\ee
First, let us calculate
\be
\star {\cal F}_5 \sim g_s M^2 dx^0\wedge dx^1\wedge dx^2\wedge dx^3
\wedge d\tau {\ell(\tau)\over K^2 h^2 \sinh^2 (\tau)}\ .
\ee 

To write down the first equation we need 
\bea
\star F_3 = M h^{-1} dx^0\wedge dx^1\wedge dx^2\wedge dx^3
\wedge 
\Bigg[ (1-F) \tanh^2 \left({\tau\over2}\right) d\tau\wedge g^1\wedge g^2&
\nonumber \\ 
+ F \coth^2 \left({\tau\over2}\right) d\tau\wedge g^3\wedge g^4
+ F' g^5\wedge (g^1\wedge g^3 +& g^2\wedge g^4)\Bigg]\ .
\eea
Assuming a constant $\phi$ and
using (\ref{nif}) we find
\be \label{mixing}
(1-F) \tanh^2 (\tau/2) - F \coth^2 (\tau/2) + 2 h {d\over d\tau}
(h^{-1} F') = \alpha (k - f) {\ell\over K^2 h \sinh^2 \tau}
\ ,
\ee
where $\alpha$ is a normalization factor
proportional to $(g_s M)^2$.

Let us turn to the second of the equations (\ref{threef}).
Since
\bea
\star H_3 =- g_s M h^{-1}
dx^0\wedge dx^1\wedge dx^2\wedge dx^3\wedge \Bigg[g^5
\wedge ( k'\tanh^2 \left({\tau\over2}\right) g^1\wedge g^2&
\nonumber \\
+ f'\coth^2 \left({\tau\over2}\right)  g^3\wedge g^4 ) 
- {1\over 2}(f-k)
d\tau\wedge (g^1\wedge g^3 + g^2\wedge g^4)\Bigg]&
\ ,
\eea
we find
\bea \label{NSt}
h {d\over d\tau} (h^{-1} \coth^2 (\tau/2) f') - {1\over 2}(f-k)= \alpha
{\ell(1-F)\over K^2 h \sinh^2 \tau} \nonumber \\
h {d\over d\tau} (h^{-1} \tanh^2 (\tau/2) k') + {1\over 2}(f-k)= \alpha 
{\ell F\over K^2 h \sinh^2 \tau}
\ .
\eea
where $\alpha$ is the same normalization factor as in (\ref{mixing}).

We have been assuming that the dilaton is constant.
The equation that guarantees this is (\ref{dilcon}).
Writing it out with our ansatz gives
\bea \label{dil}
{(k')^2\over \cosh^4 (\tau/2)}
+{(f')^2\over \sinh^4 (\tau/2)} + {2(f-k)^2\over \sinh^2 \tau}
\nonumber \\
=
{(1-F)^2\over \cosh^4 (\tau/2)}
+{F^2\over \sinh^4 (\tau/2)} + {8(F')^2\over \sinh^2 \tau}
\ .
\eea

In order to complete the system of equations we need the Einstein
equations for the metric. In view of the simplified ansatz (\ref{specans})
for the metric it is sufficient to use the trace of the Einstein
equation:
\be \label{crucial}
h^{-3/2}\nabla^2_6 h \sim g_s^2 F_3^2 + H_3^2 = 2 g_s^2 F_3^2
\ ,
\ee
where now $\nabla^2_6$ is the Laplacian on the deformed conifold.
Using (\ref{metricd}), we find that the explicit form of
this equation is
\be  {1\over \sinh^2\tau} {d\over d\tau} 
(h' K^2(\tau) \sinh^2 \tau ) = -{\alpha\over 4}\left [
{(1-F)^2\over \cosh^4 (\tau/2)}
+{F^2\over \sinh^4 (\tau/2)} + {8(F')^2\over \sinh^2 \tau} \right ]
\ .\ee

\subsection{The First-Order Equations and Their Solution}

In searching for BPS saturated
supergravity backgrounds, the second order equations should be replaced by
a system of first-order ones (see, for instance,
\cite{Freed,Gir}). Luckily, this is possible for our ansatz.
We have been able to find a system of simple first-order equations,
from which (\ref{mixing}), (\ref{NSt}), (\ref{dil}) and
(\ref{crucial}) follow:
\bea \label{firstorder}
f' &=& (1-F) \tanh^2 (\tau/2)\ , \nonumber \\
k' &=& F \coth^2 (\tau/2)\ , \nonumber \\
F' &=& {1\over 2} (k-f)\ , 
\eea
and
\be \label{firstgrav}
h' = - \alpha {f(1-F) + kF\over K^2 (\tau) \sinh^2 \tau}
\ .
\ee

Note that the first three of these equations,
(\ref{firstorder}), form a closed system and need to be
solved first. 
In fact, these equations imply the self-duality of the
complex 3-form with respect to the metric of the
deformed conifold: $\star_6 G_3 = i G_3$.\footnote{
We believe, according to a discussion in section 2.4, that
$G_3$ is a harmonic $(2,1)$ form on the deformed conifold.}
Inspection of these equations shows that the small $\tau$
behavior is\footnote{It is also possible to shift $f$ and $k$
by the same constant. The effect of this shift will
be considered in section 5.2.}
\be
f \sim \tau^3 \ ,\qquad k\sim \tau\ , \qquad
F \sim \tau^2
\ .
\ee
On the other hand, for large $\tau$ the 3-forms have to match onto
the conifold solution \cite{KT},
\be f\rightarrow {\tau\over 2}\ , \qquad
k\rightarrow {\tau\over 2}\ , \qquad
F\rightarrow {1\over 2}\ .
\ee
Remarkably, it is possible to find the solution with these boundary 
conditions in closed form.
Combining \eref{firstorder} we find the following second-order equation
for $F$:
\be
F'' = {1\over 2}[F \coth^2 (\tau/2) + (F-1) \tanh^2 (\tau/2)]
\ .\ee
The solution is 
\be
F(\tau) = {\sinh \tau -\tau\over 2\sinh\tau}\ ,
\ee
from which we obtain
\bea
f(\tau) &=& {\tau\coth\tau - 1\over 2\sinh\tau}(\cosh\tau-1) \ ,
\nonumber \\ 
k(\tau) &=& {\tau\coth\tau - 1\over 2\sinh\tau}(\cosh\tau+1)
\ .
\eea

Now that we have solved for the 3-forms on the deformed conifold,
the warp factor may be determined by integrating
(\ref{firstgrav}).
First we note that
\be
\ell(\tau) = f(1-F) + kF =  {\tau\coth\tau - 1\over 4\sinh^2\tau}
(\sinh 2\tau-2\tau)
\ .\ee
This behaves as $\tau^3$ for small $\tau$.
It follows that, for small $\tau$,
\be
h= a_0 + a_1 \tau^2 + \ldots
\ .
\ee
For large $\tau$ we impose, as usual, the boundary condition that
$h$ vanishes. The resulting integral expression for $h$ is
\be \label{intsol}
h(\tau) = \alpha { 2^{2/3}\over 4}
\int_\tau^\infty d x {x\coth x-1\over \sinh^2 x} (\sinh (2x) - 2x)^{1/3}
\ .
\ee 
We have not succeeded in evaluating this in terms of elementary
or well-known special functions.
For our purposes it is enough to show that 
\be
h(\tau\to 0) \to a_0 \ ; \qquad 
\ h(\tau\to\infty)\to {3\over 4} 2^{1/3} \alpha \tau e^{-4\tau/3}
\ .\ee
This is nonsingular at the tip of the deformed conifold and, from
\eref{changeofc}, matches the form of the large-$\tau$ solution
\eref{UVs}.  The small $\tau$ behavior follows from the
convergence of the integral \eref{intsol}, while at large
$\tau$ the integrand becomes $\sim xe^{-4x/3}$.

Thus,
for small $\tau$ the ten-dimensional geometry is 
approximately $R^{3,1}$ times the 
deformed conifold:
\be \label{apex}
ds_{10}^2 \rightarrow a_0^{-1/2} dx_n dx_n
+ a_0^{1/2} \left({1\over 2} d\tau^2  + d\Omega_3^2 + {1\over 4}\tau^2
[(g^1)^2 + (g^2)^2]\right)
\ .
\ee

Very importantly, for large $g_s M$
the curvatures found in our solution are small
everywhere.
 This is true even far in the IR. Indeed, since the integral
(\ref{intsol}) converges, 
\be
a_0\sim \alpha\sim (g_s M)^2
\ .
\ee
Therefore, the radius-squared of the $S^3$ at $\tau=0$ is of order 
$g_s M$,
which is the `t Hooft coupling of the gauge theory found far in the IR.
As long as this is large, the curvatures are small and the SUGRA
approximation is reliable.

We have now seen that the deformation of the conifold allows the
solution to be non-singular. Qualitatively, this is because the
conserved $F_3$ flux prevents the 3-cycle from collapsing. This is why
we expect to find a metric with a collapsing 2-cycle but finite
3-cycle, and these are the properties of the deformed conifold.

It may be of further interest to consider more general metrics of the
form (\ref{genans}), and to allow the dilaton to vary. In that event
it still seems likely that the qualitative properties of the solution
near the apex will not change.

\subsection{Correspondence with the Gauge Theory}

In this section we point out some interesting features of the SUGRA
background we have found and show how they realize the expected
phenomena in the dual field theory.  In particular, we will now
demonstrate that there is confinement and magnetic screening, and
argue that there are domain walls and baryon vertices with a
definite mass scale.  In many ways our results resemble those found in
the $\NN=1^*$ theory \cite{PS}, but the specific details are quite
different; the confining vacua of $\NN=1^*$ involve a spacetime
with a spherical 5-brane sitting in it, while our present spacetime
is purely given by supergravity.

First we should ask the question: how does the dimensional transmutation
manifest itself in supergravity? The answer is related to the presence of
parameter $\epsilon$ in the deformed conifold metric (\ref{metricd}).
Reinstating this parameter is accomplished through
\be
ds_6^2 \to \epsilon^{4/3} ds_6^2\ .
\ee
We are then free to redefine $h\to h \epsilon^{-8/3}$
to remove the $\epsilon$ dependence from the transverse part of the 
metric. Very importantly, the dependence then appears in the longitudinal 
part, and the metric assumes the form
\be \label{newmet}
ds^2_{10} =   h^{-1/2}(\tau)  m^2 dx_n dx_n 
 +  h^{1/2}(\tau) ds_6^2 \ ,
\ee
so that $m\sim \epsilon^{2/3}$ sets the 4-d mass scale.
This scale then appears in all 4-d dimensionful quantities.

Now let us see the theory has confining flux tubes.  The key point is
that in the metric for small $\tau$ (\ref{apex}) the function
multiplying $dx_n dx_n$ approaches a constant. This should be
contrasted with the $AdS_5$ metric where this function vanishes at the
horizon, or with the singular metric of \cite{KT} where it blows up.
Consider a Wilson contour positioned at fixed $\tau$, and
calculate the expectation value of the Wilson loop using the
prescription \cite{Juan,Rey}. The minimal area surface bounded by the
contour bends towards smaller $\tau$. If the contour has a very large
area $A$, then most of the minimal surface will drift down into the
region near $\tau=0$. From the fact that the coefficient of $dx_n
dx_n$ is finite at $\tau=0$, we find that a fundamental string with
this surface will have a finite tension, and so the resulting Wilson
loop satisfies the area law.  Since for large $g_s M$ the SUGRA
description is reliable for all $\tau$, we seem to have found a ``pure
supergravity proof'' of confinement in ${\cal N}=1$ gauge theory.  A
similar result was found previously in \cite{PS} but involved a
spacetime containing an NS5-brane with D3-brane charge.  A simple
estimate shows that the string tension scales as 
\be 
T_s \sim
{m^2\over g_s M} \ .  
\ee

To see that magnetic charge is screened, we must identify the correct
massive magnetically-charged source.  The correct choice is a
fractional D1-brane, that is, a D3-brane wrapped on the $S^2$ of
$T^{11}$, attached to the boundary of the space at $\tau=\infty$.  On
the six-dimensional deformed conifold the $S^2$ is fibered over $\tau$
such that the resulting three-dimensional bundle has only one boundary,
at $\tau=\infty$; near $\tau=0$ the $S^2$ shrinks to zero size
and the bundle locally has topology $R^3$.  Therefore, a
D3-brane with a single boundary can be wrapped on this bundle,
corresponding to a fractional D1-brane attached at $\tau=\infty$ which
quietly ends at $\tau=0$.  Strictly speaking, this only shows monopole
charge is not confined; to show it is screened one must go further and
show this object does not couple to any massless modes.

  As we showed in section 2, the field theory has an anomaly-free
$\ZZ_{2M}$ R-symmetry at all scales. The UV limit of our background,
which coincides with the solution found in \cite{KT}, has a $U(1)$
R-symmetry associated with the rotations of the angular coordinate
$\psi$. For large $M$ it is is somewhat difficult to distinguish
between the $U(1)$ and its discrete subgroup $\ZZ_{2M}$. In fact, the
anomaly in the $U(1)$, which breaks it down to $\ZZ_{2M}$, is an
effect of fractional D-instantons, the euclidean D-string world sheets
propagating inside $T^{11}$. The Wess-Zumino term present in the
D-string action, which is associated with the topologically
non-trivial $F_3$, has to be quantized (this is simply the $F_3$ flux
quantization). As a result, the phase in the D-string path integral
assumes $\ZZ_{2M}$ rather than $U(1)$ values.

Our metric provides a geometrical realization for the phenomenon of
chiral symmetry breaking found in the field theory; the dynamical
breaking of the $\ZZ_{2M}$ down to $\ZZ_2$ occurs via the deformation
of the conifold. In the pure supergravity limit we have discussed, the
spontaneous chiral symmetry breaking generates an $\eta'$-like
Goldstone boson (the zero mode in our solution corresponding to
rotation of the coordinate $\psi$), which must get a mass of order
$1/M$ from these fractional instantons.  To see how this mass arises,
and how it relates to the domain walls which we discuss in a moment,
would be very interesting.

It is by now clear why the conifold ansatz adopted in \cite{KT} and reviewed
in section 2 is too restrictive: it has the $U(1)$ symmetry everywhere.
On the other hand, our deformed conifold ansatz breaks it down to
$\ZZ_2$, with the $U(1)$ symmetry becoming asymptotically restored 
at large radius. Thus, the deformation of the conifold ties together
several crucial IR effects: resolution of the naked singularity found
in \cite{KT}, breaking of the chiral symmetry down to $\ZZ_2$, and
quark confinement. At the same time, the deformation does not
destroy the logarithmic running of the couplings found in \cite{KT}
because it does not affect the geometry far in the UV.

Due to the deformation, the full SUGRA background has a finite
3-cycle.  We now interpret various branes wrapped over this 3-cycle in
terms of the gauge theory. Note that the 3-cycle has the minimal
volume near $\tau=0$, hence all the wrapped branes will be localized
there.  This should be contrasted with wrapped branes in $AdS_5\times
X_5$ where they are allowed to have an arbitrary radial coordinate.  A
wrapped D3-brane plays the role of a baryon vertex which ties
together $M$ fundamental strings.  Note that for $M=0$ the D3-brane
wrapped on the $S^3$ gave a dibaryon \cite{GK}; the connection between
these two objects becomes clearer when one notes that for $M>0$ the
dibaryon has $M$ uncontracted indices, and therefore joins $M$
external charges.  Meanwhile, a D5-brane wrapped over the $S^3$
appears to play the role of the domain wall separating two
inequivalent vacua of the gauge theory.  As we expect, flux tubes can
end on this object \cite{dvalishif}, and baryons can dissolve in it; as in
\cite{PS}, we may also build the domain walls from the baryons.
Indeed, D3 and D5-branes play the roles of baryon vertices and
domain walls in $\NN=1^*$; however in that case they do not wrap a
cycle but instead have a boundary on the NS5-brane in the space
\cite{PS}. Calculations using the metric (\ref{newmet}) show that the
baryon mass is
\be
M_b \sim m M\ ,
\ee
while the D5-brane domain wall tension is
\be
T_{wall} \sim {1\over g_s} m^3\ .
\ee 

Additionally, one can obtain the glueball spectrum in this theory.  To
do so requires finding the spectrum of eigenmodes of various
supergravity fields in the metric background we have constructed.
Since the background is known explicitly as a function of $\tau$,
the calculation should be no more difficult than in \cite{glueballucb,
glueballbrown}.  Unlike the case of $\NN=1^*$, where the
presence of a narrow throat near a single NS5-brane could make the
computation potentially unreliable for the lowest modes \cite{PS},
there is no possible subtlety here, as the bulk space is large and
everywhere nonsingular.  Of course, there will be Kaluza-Klein modes
on the $S^3$ which are not present in the pure \none\ Yang-Mills
theory.  These are analogous to the extra modes which appear in both
\cite{Wittenthermo} and \cite{PS}; their presence is expected, since they
are necessary whenever pure \none\ Yang-Mills is embedded into a
theory that is fully in the supergravity regime.  Only in the limit of
pure \none\ Yang-Mills, which we discuss below, can
they be removed. A simple estimate of the glueball and KK modes masses
shows that, in the SUGRA limit both scale as $m/(g_s M)$.
Comparing with the string tension, we see that
\be 
T_s \sim g_s M (m_{glueball} )^2
\ .\ee
Thus, there is a large separation of scales between string tension
and glueball mass in supergravity (a similar problem was observed in
\cite{glueballucb,glueballbrown}) which goes away at small $g_sM$.

Finally, we should address the possibility that $N$ is not a multiple of $M$.
Note that in our solution the 5-form flux vanishes for $\tau=0$:
\be
\int F_5 = \ell(\tau)\sim \tau^3
\ .
\ee
This suggests that the IR solution given above describes a large number
$M$ of wrapped D5-branes without any D3-branes. Therefore, for small
$\tau$ the background should be dual to $SU(M)$ gauge theory (the
SUGRA is reliable only if both $M$ and $g_s M$ are large).  More
generally, however, the field theory analysis tells us that theories
that may arise in the IR have gauge groups $SU(M+p)\times SU(p)$, with
$M>p\geq 0$.  If $M$ is large and $p$ is of order $1$, then the dual
supergravity background should be the same as for $p=0$, to leading
order in $M$. The extra $p$ colors should come from $p$ actual
D3-branes, placed at various points in our background. Then the moduli
space for each D3-brane is essentially the deformed conifold, in
agreement with the field theory analysis.  When far from $\tau=0$, the
D3-branes represent the IR \nfour\ $SU(p)$ factor in the theory.  The
't Hooft
coupling on these branes is $g_s p\ll 1$, so when they are brought to
$\tau=0$ the theory represented is essentially $SU(M+p)$ with $2p$
classically massless flavors and a quartic superpotential.  The
nonperturbative analysis of this theory, given in section 4 and in the
appendix, then applies, giving chiral symmetry breaking and a moduli
space with $M$ branches.

Note that confinement is lost in the presence of the D3-branes, in
agreement with the field theory.  Strings hanging from the boundary
can simply end on the D3-branes, corresponding to the statement that
external sources are screened by massive dynamical quarks and end up
carrying only $SU(p)$ charge.\footnote{Similar findings were also
obtained in $\NN=1^*$ \cite{PS}.  Many of the $\NN=1^*$ vacua have
dynamical massive $W$-bosons, whose pair production eliminates
confinement.  The representation of this gauge theory physics in the
string theory is closely related to the representation presented here
and in the last paragraph of this section.}  The corresponding Wilson
loop will have a perimeter law. Of course if the quarks are heavy
({\it i.e.}, if the D3-branes are at large $\tau$) then relatively
short flux tubes should be stable.  It would be interesting to
actually demonstrate this fact, which follows not from topology but
from quantum dynamics.

On the other hand, if $p$ is of the same order as $M$, then the flux
due to the D3-branes is large and should be included in the SUGRA
solution. First, let us try to change the boundary condition
on $F_5$ so that  
$F_5$ no longer vanishes at
$\tau=0$ but is $\sim p$. We find a consistent
solution for the 5-form
by replacing $\ell(\tau)\to \ell (\tau) + C$, where $C$ is a
constant of order $p/(g_s M^2)$.  From (\ref{firstgrav}) 
we find that the effect of this on the warp factor is
$h\to h+\tilde h$ where
\be
\tilde h(\tau)=\alpha C \int_\tau^\infty dx
{1\over K^2 (x) \sinh^2 x}
\ .
\ee
This yields a
singular behavior of $\tilde h$ for small $\tau$:
\be \label{newbeh}
\tilde h\sim {\alpha C\over \tau}
\ .
\ee

The new behavior of $h$ does change significantly the physical
interpretation of the solution.  Now the coefficient of the $dx_n
dx_n$ term scales as $\tau^{1/2}$ for small $\tau$; hence, the Wilson
loop no longer satisfies the area law.  Again, we find agreement with
the field theory.  This gravity background corresponds to making the
charged matter as light as possible (that is, making the expectation
values of the scalar fields all as small as possible.)  In this regime
we expect no metastable flux tubes; the dynamical charges in the
fundamental representation of $SU(M+p)$ will screen external electric
sources, until the sources are charged only under $SU(p)$, which does
not confine.  

Thus, the new behavior of the metric (\ref{newbeh}) incorporates the
loss of confinement found upon addition of dynamical quarks.
However, supergravity may receive large corrections in the small
$\tau$ region because curvatures blow up at $\tau=0$
where we find a singular horizon.\footnote{
We are grateful to A. Tseytlin for useful discussions
of this point.} Thus, requiring that $F_5$ does not vanish
at $\tau=0$ actually causes a singularity.
Can we construct a non-singular SUGRA solution which incorporates
screening? We believe that the correct approach may be to add
D3-brane sources with total charge $p$ (this way $F_5$ may
smoothly turn on from zero at $\tau=0$ to $p$ at some finite value of
$\tau$). This idea also agrees with the incorporation of small
$p$ via actual D3-branes. We postpone construction of such non-singular
`Coulomb branch' solutions until a later publication.

\subsection{The Dual of Pure ${\cal N}=1$ Yang-Mills Theory}

As we have shown above, supergravity serves as a reliable dual of a
cascading $SU(N+M)\times SU(M)$ gauge theory, provided that $g_s M$ is
very large. We have also shown that, under appropriate circumstances,
at the bottom of the cascade, we essentially have an $SU(M)$ theory,
with the other gauge group disappearing.  An immediate
question that arises is: can our results be used to learn something
about the pure glue ${\cal N}=1$ theory?

To start answering this question, let us note that the field $B_2$ is
multiplied by $g_s M$, while the jumps in the cascade occur after
$B_2$ has changed by an amount of order $1$.  Thus, the range of
$\tau$ which describes any particular gauge group in the cascade is of
order $1/(g_s M)$.  This implies the supergravity regime is not
sufficient for constructing such a dual, because for large $g_s M$ the
cascade jumps occur very frequently, and we find the pure glue theory
only for small $\tau$.  There, at the tip of the deformed conifold,
both $B_2$ and $F_5$ are very small, $F_3$ is of order $M$, and the
metric is approximately given by (\ref{apex}). 

To have a reliable dual of the pure glue theory, valid for a large
range of $\tau$, we need to take the limit of {\it small} $g_s M$ (and
thus small $B_2$, holding $M$ fixed) which is the opposite of the limit
where supergravity has small corrections.  In this limit the $S_3$ at
the apex of the conifold becomes small and the space acquires large
curvature. This situation is familiar from previous studies aimed at
finding a string theory dual of a pure glue gauge theory
\cite{Wittenthermo,PS}.

Nevertheless, our work does constitute progress towards formulating a
stringy dual because our SUGRA background captures the correct
topology of the resulting string background.  Indeed we are led to
conjecture that the type IIB string dual of the pure glue ${\cal N}=1$
$SU(M)$ theory is given by a $g_s M \to 0$ limit of a warped
deformed conifold background, with $M$ units of the $F_3$ flux
piercing its 3-cycle, and with $B_2$ and $F_5$ approaching zero
at the apex. This would
be the space generated by the fractional D3-branes alone, with {\it
no} admixture of regular D3-branes. Hence it is relevant
to pure $SU(M)$ theory with no quark flavors.
Of course, studying such a theory
for small $g_s M$ is difficult due to the well-known problems with RR
flux and large curvature.  However, the self-dual 5-form flux, which
brings in some additional problems, is small, which raises hopes of a
novel sigma model formulation.

We note also that the addition of a small number of D3-branes to this
story will permit the study of the $SU(M+p)\times SU(p)$ theory, which
essentially reduces, for small $g_s$ and $p\ll M$, to $SU(M+p)$ with
$2p$ flavors and an all-important quartic superpotential.  It is far
from certain that this construction can give any insight into QCD,
since the light charged scalars play such a central role in the
dynamics.  However, if these scalars can easily be removed (along with
the gauginos) through explicit supersymmetry breaking, there might be
additional interest in this approach.

\section{Discussion}

We have not addressed the question of how to compute field theory
correlation functions in this context, where our space does not
approach Anti-de-Sitter space at large $r$.  However, it is easy to
see this space still has a boundary, and from the behavior of
$h(\tau)$ it is clear that the logarithm is a subleading effect at
large $r$.  Correspondingly, at large $N\gg M$, there is a sense in
which the operators in the field theory have the same spectrum that
they have for $M=0$, since $\gamma\approx-\half$. We therefore believe
that for low-lying supergravity modes, corresponding to operators of
dimension much less than $M$, the story will not be modified in a
significant way from that discussed in \cite{US,EW}.  For operators of
dimension $\Delta>{3\over 2}M$, we expect more interesting effects.
These operators appear to exist at scales where ${3\over 2}N>\Delta$,
but should be eliminated when ${3\over 2}N<\Delta.$ In the gauge
theory, it is known what should occur \cite{KSS}; operators of high
dimension present classically are actually removed by quantum effects,
which in the low-energy dual theory appear as simple group theory.  On
the gravity side we may speculate that high-lying bulk modes propagate
in from the boundary until the region where $N\sim {3\over 2}\Delta$; there
$T^{11}$ has shrunk down such that these modes blow up into the
``giant gravitons'' of \cite{giant}.  Only modes with dimension
$\Delta < {3\over 2}M$ can propagate all the way to $\tau=0$.

It is easy to see that our story of the duality cascade can be
orientifolded.  This is obvious from the type IIA string theory brane
construction.  It is also clear from the corresponding $SO\times Sp$
gauge theory, although we have not analyzed the field theory dynamics
to see how the orientifolded conifold is deformed.  A number of other
modifications, including theories whose IIA version involves multiple
NS and NS' branes, could potentially be interesting.  This might be
especially true for theories which are qualitatively different in the
infrared from pure Yang-Mills, such as those studied in \cite{KS}.

Another interesting choice would be to orbifold the theory along the
lines of \cite{PS}, so that the low energy theory is
non-supersymmetric $SU(M)^2$ with a Dirac fermion in the bifundamental
representation.  In contrast to the case studied in \cite{PS}, the
masslessness of this fermion would be exact, as it is guaranteed by
the $\ZZ_{2M}$ R-symmetry, and therefore chiral symmetry breaking and
confinement in this QCD-like theory could be exhibited in the supergravity
regime.

Finally, it is interesting to resurrect a scenario discarded five
years ago for its apparent absurdity.  Namely, it is conceivable that
the standard model --- a small gauge group --- itself lies at
the base of a duality cascade.  This is certainly possible, since the
addition of supersymmetry and some appropriately chosen massive matter
at the TeV scale easily could make the theory into one which could
emerge from such an RG flow.  In \cite{dualitywall} it was in fact
pointed out that this was the natural scenario if the standard model,
with its very small gauge groups, is a low-energy Seiberg-dual
description of some other theory; every natural choice for an
ultraviolet theory has a larger gauge group than the standard model,
and typically hits a Landau pole below the Planck scale, requiring
additional duality transformations, still larger gauge groups, more
Landau poles, and continuation {\it ad nauseum}.  This was termed the
``duality wall'' (since in some cases the duality transformations
piled up so fast that an infinite number were required in a finite
energy range.)  But now we see this continuous generation of larger
and larger gauge groups --- ugly and unmotivated within field theory,
and driving the field theory into highly non-perturbative regimes ---
can correspond to a perfectly natural spacetime background on which
strings may propagate.  If we imagine that the ultraviolet of the
duality cascade is cut off in a compact space (along the lines of
\cite{Verlinde}, following \cite{RShier}) we may conjecture that the
standard model coupled to gravity is best described, at high energy,
by a compactified string theory on a space with a logarithmic (or
otherwise warped) throat, with the weakly coupled standard model
emerging as a good description only at energies below, say, 1--100
TeV.  Such a model provides another possible way, somewhat related to
ideas of \cite{Verlinde,RShier,FramVafa,GoldWise,Polpriv}, to explain
the hierarchy between the gravitational and electroweak scales: it is
perhaps given by TeV$= m_{Pl}\times e^{-c N/M}$, where $M$ is of order
2 to 5, $c$ is a number of order one, and $N$ is the number of colors
of the gauge group at the Planck scale.

\section*{Acknowledgements}
We are grateful to K. Dasgupta, 
S. Frolov, S. Gubser, S. Gukov, J. Maldacena,
J. Polchinski, A. Tseytlin and E. Witten for useful discussions.  The
work of I.K.  was supported in part by the NSF grant PHY-9802484 and
by the James S. McDonnell Foundation Grant No. 91-48; that of
M.J.S. was supported by NSF grant PHY95-13835 and by the W.M. Keck
Foundation.

\section{Appendix}
In this appendix we analyze the field theory in somewhat
greater detail, confirming and extending the results of section 3.

First, we may check the results of section 3 in
another region of moduli space.  Consider first $SU(M+1)$ with two
flavors.  Suppose we permit $C_1$ and $D_1$ to have equal
expectation values $v$, so that $N_{11}=v^2$. This breaks the
$SU(M+1)$ to $SU(M)$.  If $\lambda$ were zero, this would leave
$SU(M)$ with one flavor $C_2$ and $D_2$, plus two gauge singlets
$\vev{C_1} D_2= N_{12}$ and $C_2\vev{D_1}=N_{21}$; the corresponding
strong coupling scale would be $\Lambda^{M+1}/v^2$.  However, the
presence of nonzero $\lambda$ gives mass to these fields, leaving the
$SU(M)$ gauge theory with a flavor of mass $\lambda v^2$.  The
effective Lagrangian is then
\begin{equation}
W = 2\lambda v^2 N_{22} + \left[\Lambda^{M+1}/v^2\over
N_{22}\right]^{{1\over M -1}}
\end{equation}
which again leads to $M$ branches with the correct values of
$N_{ij}N_{k\ell} \eps^{ik}\eps^{j\ell}$.  

That our discussion of the $SU(M+1)$ theory in section 3 was only part
of the story can be seen by starting one step higher, with
$SU(2M+1)\times SU(M+1)$, which reduces after one duality
transformation to the $SU(M+1)$ case.  The $SU(2M+1)$ gauge group has
one more flavor than color, and therefore, as $\lambda_1\to 0$, the
theory is governed by the results of \cite{nsexact}.  For
$\lambda_1=0$ the superpotential must go over to
\begin{equation}
W \to {\det P_{ijb}^a\over\Lambda_1^{4M+1}} - C_{ia}P_{ijb}^a D_{j}^b \ ,
\end{equation}
 where $a,b$ are color indices of $SU(M+1)$, and $P\sim AB$, $C\sim
A^{2M+1}$, $D\sim B^{2M+1}$.  From this we learn the function
$F_1(I_1,J_1,R_1^{(1)})$ is not equal to one, and in fact, in the limit
$\lambda_1\to 0$, that is, $I_1,J_1\to 0$, we have
$F_1(I_1,J_1,R_1^{(1)})\to \sqrt{I_1/J_1}f(R_1^{(1)})$.  The low energy
theory is then $SU(M+1)$ with two flavors $C_i,D_i$ but with
superpotential
\begin{equation}
W = \lambda_2 C_iD_jC_kD_l\eps^{ik}\eps^{jl}F_2(I_2,J_2)
\end{equation}
Here $(C_iD_jC_kD_l\eps^{ik}\eps^{jl})$ is
the only invariant involving $C$ and $D$; there are no $R_2$ ratios.
The low energy effective superpotential is now
\begin{equation}
W_L = \lambda N_{ij}N_{k\ell} \eps^{ik}\eps^{j\ell}F_2(I_2,J_2) .
\end{equation}
Note that $F_2(I_2\to 0,J_2\to 0) = 1 + \sqrt{I_2/J_2}$; some other
limits can be studied but will not be needed here.  The vacuum
equations are
\begin{equation}
0 = \lambda \left[F(I_2,J_2) + I_2 
{\partial F(I_2,J_2)\over\partial I_2}\right] N_{ij}.
\end{equation}
This gives an equation for $I_2 \propto 
(N_{ij}N_{k\ell} \eps^{ik}\eps^{j\ell})^{2M}$, whose solution must be
\begin{equation}
I_2 = \ G(J_2) \ .
\end{equation}
The holomorphic function $G(J_2)$ is not zero everywhere (since for
$I_2\to 0$, $J_2\to 0$ it is not zero) so it can only be zero at
special points.  Consequently $N_{ij}N_{k\ell} \eps^{ik}\eps^{j\ell}$
is generally nonzero.  Since the $\ZZ_{2M}$ symmetry rotates
$N_{ij}N_{k\ell} \eps^{ik}\eps^{j\ell}$ by $e^{2\pi i/M}$, we again
find $M$ separate branches.  Again there are no restrictions on
the individual values of the $N_{ij}$, so each branch takes the form of a
deformed conifold, with a nonzero superpotential.  Thus we obtain the
same result as before; only the magnitude of the deformation is
modified from our previous analysis.

This analysis is too weak to rule out the possibility
that there might be several independent solutions for $I_2$ given
a single value of $J_2$.  This would lead to several sets of branches,
each set consisting of $M$ copies of a deformed conifold;
the different sets would have deformations of different magnitudes.
In the limit $\Lambda_{1}\to \infty$ only one set would remain,
as in our earlier analysis of the $SU(M+1)$ theory.

Next, we consider the case of $SU(M+p)\times SU(p)$, 
$0<p<M$. 
We will
first perform the analysis by taking the $SU(p)$ coupling small.  We
define $(N_{ij})^\alpha_\beta = (C_i)^\alpha_a (D_i)^a_\beta$, where
$\alpha,\beta$ are $SU(p)$ indices and $a,b$ are $SU(M+p)$ indices.
If the $SU(p)$ coupling were set to zero, then we would have an
$SU(M+p)$ gauge theory with $2p$ flavors.  An Affleck-Dine-Seiberg
superpotential would be generated, giving
\begin{equation}
W = \lambda (N_{ij})^\alpha_\beta (N_{k\ell})_\alpha^\beta
 \eps^{ik}\eps^{j\ell} + (M-p) \left({\Lambda_{1}^{3M+p}\over
 \det_{ij,\alpha\beta} N }\right)^{1\over M-p} \
\end{equation}
where in the determinant we treat $N$ as a $2p\times 2p$ matrix.
A little algebra gives the equations for a supersymmetric vacuum as
\begin{equation}
\det[(N_{ij})^\alpha_\beta] \propto
\left({\Lambda_{1}^{3M+p}\over \lambda^{M-p}
}\right)^{2\over M} 
\end{equation}
and
\begin{equation}
\lambda (N_{ij})^\alpha_\beta (N_{k\ell})_\alpha^\beta \eps^{ik}\eps^{j\ell}
\propto (\lambda^p\Lambda^{3M+p})^{1/M}
\end{equation}
It is possible to show that these equations represent $M$ branches,
each of which is $p$ copies of
the deformed conifold --- in other words, the moduli space of $p$ probe
branes moving on the deformed conifold.  First, note that
$N^0_{ij}\equiv (N_{ij})^\alpha_\alpha$ is a gauge invariant operator.
If we demand that the $SU(p)$-adjoint fields
$(N_{ij})^\alpha_\beta - {1\over p}
\delta^\alpha_\beta(N_{ij})^\gamma_\gamma$ vanish, then the equations
above become
\begin{equation}
\det[(N^0_{ij})] \propto\left({\Lambda_{1}^{3M+p}\over \lambda^{M-p}
}\right)^{2\over M} 
\end{equation}
and
\begin{equation}
\lambda N^0_{ij}N^0_{k\ell} \eps^{ik}\eps^{j\ell}
\propto (\lambda^p\Lambda^{3M+p})^{1/M}
\end{equation}
which gives $M$ branches, each of which is a single copy of the
deformed conifold.  This region of moduli space corresponds to taking
all $p$ probe branes to have the same positions on the conifold.  As
before the $\ZZ_{2M}$ global symmetry is broken to $\ZZ_2$; it is easy
to see that the superpotential on the $M$ branches rotates by a phase
under the broken $\ZZ_M$.  Expectation values for elements of
$(N_{ij})^\alpha_\beta - {1\over p}
\delta^\alpha_\beta(N_{ij})^\gamma_\gamma$ correspond to moving the
$p$ probe branes apart; taking the special cases where these fields
are diagonal, it is easy to show that each set of eigenvalues of
$(N_{ij})^\alpha_\beta$, $i,j=1,2$, sweeps out its own copy of the
deformed conifold.

When the $SU(p)$ gauge coupling is turned back on, the superpotential
will include unknown functions of the invariants $I$, $J$, and $R$.
These can be generated by a number of different physical phenomena,
including instantons in regions where the $SU(p)$ group is partially
broken.  However, as before, these functions change the quantitative
features of the deformation of the conifold without altering the basic
picture we have obtained.  Furthermore, we expect no additional
significant infrared dynamics. Above the strong-dynamics scale for
$SU(M+p)$, the $SU(p)$ gauge group is infrared free.  Below it, the
$SU(p)$ group contains three adjoint fields $(N_{ij})^\alpha_\beta$
which have a trilinear superpotential --- in short, a copy of \nfour\
Yang-Mills.  The $SU(p)$ sector is therefore scale-invariant and
nonconfining at low energy.  Lastly, we expect that the $SU(p)$
dynamics plays no role in the supergravity regime for $p\ll M$.
Supergravity requires we work at small gauge coupling and large 't
Hooft coupling for $SU(M)$, but in this regime $SU(p)$ will have small
't Hooft coupling and will be described by weakly-coupled field
theory. In the end, then, we again expect $M$ branches, given by
equations of the same qualitative form as above.

The case $p=M$ is the most subtle. For $SU(2M)\times SU(M)$,
the $SU(2M)$ theory has equal numbers of flavors and colors,
and consequently its moduli space is modified quantum 
mechanically \cite{nsexact}.  If we turn off the $SU(M)$ coupling,
the superpotential becomes
\begin{equation}
W = \lambda (N_{ij})^\alpha_\beta
(N_{k\ell})_\alpha^\beta\eps^{ik}\eps^{j\ell}F_1(I_1/J_1)
+ X(\det [(N_{ij})^\alpha_\beta]
-{\cal B}\bar{\cal B} - \Lambda_{2M}^{4M}) \ ,
\end{equation}
where the ``baryon'' ${\cal B}$ is the gauge invariant operator
$A_1^MA_2^M$, and the anti-baryon is similarly constructed from $B_i$.
Here the equations seem to have multiple solutions.  One solution is
\begin{equation}
X = 0 \ ; \ N = 0 \ ; {\cal B} = \bar{\cal B} = i \Lambda_{2M}^{2M} \ .
\end{equation}
In this case, the $SU(M)$ gauge group is unbroken and, when its
coupling is restored, it generates $M$ distinct and isolated vacua via
usual gaugino condensation.  Alternatively, we may have
\begin{equation}
{\cal B} = \bar{\cal B} = 0; \det [(N_{ij})^\alpha_\beta] =  \Lambda_{2M}^{4M}
;  [(N_{ij})^\alpha_\beta
(N_{k\ell})_\alpha^\beta\eps^{ik}\eps^{j\ell}G_1(I_1/J_1)]^M 
= \Lambda_{2M}^{4M},
\end{equation}
where we have not determined $G_1$.  As before this leads to $M$
branches, each of which has $M$ probe branes moving on a deformed
conifold.

This suggests that the complete solution to a theory with gauge group
$SU(N+M)\times SU(N)$ might involve not one set of $M$ branches but
many.  The smallest set would consist of $p \equiv N\ {\rm mod}\ M$
D3-branes moving on the deformed conifold.  The next smallest set
would consist of $p+M$ D3-branes.  Next would follow a branch with
$p+2M$ D3-branes, and so forth, growing in size without limit.  To see
whether this is the case requires a more thorough and complete field
theory analysis, which we have not performed.

In any case, these partial results all support the main claims of 
the paper: that all branches which appear are consistent with
probe branes moving on a deformed conifold, and that each branch
is one of $M$ identical branches which are rotated by the
spontaneously broken $\ZZ_{2M}$ R-symmetry.


\end{document}